# Seeded x-ray free-electron laser generating radiation with laser statistical properties

O. Yu. Gorobtsov,[1] G. Mercurio,[2] F. Capotondi,[3] P. Skopintsev,[1,4] S. Lazarev,[1,5] I.A. Zaluzhnyy,[1,6] M. Danailov,[3] M. Dell`Angela,[7] M. Manfredda,[3] E. Pedersoli,[3] L. Giannessi,[3,8] M. Kiskinova,[3] K. C. Prince,[3,9] W. Wurth,[1,2] and I. A. Vartanyants[1,6,*]

[1]*Deutsches Elektronen-Synchrotron DESY, Notkestrasse 85, D-22607 Hamburg, Germany*
[2]*Department of Physics, University of Hamburg and Center for Free Electron Laser Science, Luruper Chausse 149, D-22761 Hamburg, Germany*
[3]*Elettra-Sincrotrone Trieste, 34149 Basovizza (Trieste), Italy*
[4]*Present address: Paul Scherrer Institute, PSI Aarebrucke, 5232 Villigen, Switzerland*
[5]*National Research Tomsk Polytechnic University (TPU), pr. Lenina 2a, 634028 Tomsk, Russia*
[6]*National Research Nuclear University MEPhI (Moscow Engineering Physics Institute), Kashirskoe shosse 31, 115409 Moscow, Russia*
[7]*CNR- IOM Istituto Officina dei Materiali, 34149 Trieste, Italy*
[8]*ENEA C.R. Frascati, Via E. Fermi 45, 00044 Frascati (Rome), Italy*
[9]*Molecular Model Discovery Laboratory, Department of Chemistry and Biotechnology, School of Science, Melbourne, Victoria 3122, Australia.*

(Dated: July 21, 2018)

[*]Corresponding author: ivan.vartaniants@desy.de

**The invention of optical lasers led to a revolution in the field of optics and even to the creation of completely new fields of research such as quantum optics [1]. The reason was their unique statistical and coherence properties. The newly emerging, short-wavelength free-electron lasers (FELs) [2-6] are sources of very bright coherent extreme-ultraviolet (XUV) and x-ray radiation with pulse durations on the order of femtoseconds, and are presently considered to be laser sources at these energies. Most existing FELs are highly spatially coherent [7, 8] but in spite of their name, they behave statistically as chaotic sources [9-11]. Here, we demonstrate experimentally, by combining Hanbury Brown and Twiss (HBT) interferometry [12, 13] with spectral measurements that the seeded XUV FERMI FEL-2 source [3] does indeed behave statistically as a laser. The first steps have been taken towards exploiting the first-order coherence of FELs [14,15], and the present work opens the way to quantum optics experiments [16] that strongly rely on high-order statistical properties of the radiation.**

Glauber in his pioneering work [1] stated that a truly coherent source should be coherent in all orders of intensity correlation functions. Most sources of radiation in the optical wavelength range behave statistically as thermal or chaotic sources. Optical lasers, due to their significantly different radiation properties, provide unique opportunities in science and technology. As first demonstrated in the time domain, single-mode or phase-locked optical lasers are not only coherent in the first-order, but also in the second-order of intensity correlation functions [17,18]. This is also valid in the spatial domain and distinguishes laser sources from chaotic sources of radiation. This difference is especially important for quantum optics experiments that are extended now to classical fields [16,19], in which high-order coherence properties of the source play an important role. The possibility of employing similar properties of sources in the XUV and x-ray range with extremely high brightness is the main attraction for completely new and exciting applications of FELs.

Most of the presently operating FELs generate radiation using the self-amplified spontaneous emission (SASE) process [20] where the radiation is produced stochastically by the electron bunch shot noise. As a result, the spatial and especially temporal structure of the x-ray pulse fluctuates strongly from shot to shot. Typically, each SASE pulse contains a large number of longitudinal modes without any phase locking between them. As a result, SASE FELs are highly coherent sources in first-order (with the degree of spatial coherence reaching 80% [7, 8]), but from a statistical point of view, they behave as chaotic sources of radiation [9-11]. To date, no consistent high-order statistical measurements have been performed at any seeded FEL.

FERMI is the first seeded single-pass FEL in the XUV regime and hosts two sources FEL-1 and FEL-2 [3]. Recent measurements demonstrated that FERMI FEL-1 is coherent in the temporal domain [21, 22]. The question naturally arises whether this indicates only first-order coherence, or whether the FERMI FEL is also coherent in higher orders, thus satisfying the definition of Glauber. This is an important question, as it defines the potential of this light source in quantum optics experiments. To answer this question we employed high-order correlation measurements at FERMI.

The basic idea behind HBT interferometry [12, 13] is to extract statistical properties of radiation from the normalized second-order intensity correlation function [22]

$$g^{(2)}(\boldsymbol{r}_1, \boldsymbol{r}_2) = \langle I(\boldsymbol{r}_1)I(\boldsymbol{r}_2)\rangle \big/ \langle I(\boldsymbol{r}_1)\rangle\langle I(\boldsymbol{r}_2)\rangle, \qquad (1)$$

where I($\boldsymbol{r}_1$) and I($\boldsymbol{r}_2$) are intensities at different spatial positions measured simultaneously, and the brackets < … > denote averaging over a large ensemble of different radiation pulses. The statistical behaviour of the $g^{(2)}$ correlation function is fundamental in quantum optics [18] and strongly depends on the radiation type. For example, for coherent laser sources $g^{(2)}$ is equal to one [17, 18] but for chaotic sources it behaves quite differently (see Methods equation (4)).

Measurements were performed at the DiProI end-station of FERMI using the FEL-2 source and acquiring simultaneously data from the on-line spectrometer PRESTO (see Fig. 1 and Methods). This feature of FERMI allowed us to analyse simultaneously the spectral profile of each pulse delivered and measured at the DiProI end-station. One more important feature of FERMI is that its operation can be switched from seeded to SASE mode [24].

We first analysed the x-ray intensity distribution of the FEL pulses in the seeded and SASE regimes of FERMI FEL-2 by employing HBT correlation analysis, equation (1). Intensity correlation functions measured in the vertical direction for both operation modes are shown in Fig. 2 (a,c) (for the horizontal direction see Fig. 6). The contrast values of the intensity correlation function (see Methods equation (3)) in the central part of the beam were on the order of 0.03 – 0.04 for the seeded beam and slightly higher about 0.1 – 0.15 for the SASE. The remarkable difference between the two regimes of operation becomes evident by correlating these observations to the corresponding spectral profiles.

The average spectrum, both in the seeded and SASE regimes of FERMI operation, was approximately Gaussian in shape but with a quite different relative bandwidth $\Delta\omega/\omega_0$, see Fig. 2(b,d). In the seeded regime it was about $4.6 \cdot 10^{-4}$ and in SASE $6 \cdot 10^{-3}$, an order of magnitude broader than in the seeded regime. The low values of the contrast measured in the seeded regime, in the case of a chaotic Gaussian beam, would indicate the presence of at least 25 - 30 independent longitudinal modes [9-11]. However, from our on-line spectrometer measurements, we observed from one to five modes varying from pulse to pulse, with one mode usually dominating. At the same time, the values of contrast obtained in SASE mode match the number of observed spectral modes (about 10), supporting the assumption of chaotic character of the source in the SASE regime (see Methods equation (4)).

Next, we proceeded with analysis of the spectral dependence of the $g^{(2)}$ function in the seeded regime of FERMI operation by implementing a sorting procedure (see Appendix B). Two data sets of pulses were then considered: $10^3$ pulses with the largest contribution of the main mode and $10^3$ pulses with the smallest contribution. Intensity correlation functions determined by equation (1) for these two data sets are presented in Fig. 3 (a,c) and examples of the single and multimode pulses are shown in Fig. 3 (b,d). Remarkably, these two data sets produce similar correlation functions with low values of contrast of about 0.02 and 0.07 for the data set with the smallest and largest contribution of the main mode, respectively. Based on these results, we conclude that the seeded FERMI FEL-2 source is not behaving as a chaotic source but rather as a laser source, even in the case when several modes are present in the spectrum. This implies that the spectral modes are potentially phase locked, as in the case of FERMI FEL-1 [21, 22], which is an important finding for coherent control experiments [14].

To further investigate the difference between the seeded and SASE operation modes, the probability distribution of total intensity was studied in both cases, Fig. 4 (a,c). As known from the first-order coherence theory, there is little difference between a multimodal laser without phase locking and a chaotic source [25]. In both cases the total pulse intensity distribution follows a Gamma distribution. We indeed observed such behaviour in the SASE regime of FERMI with about 8 longitudinal modes (see Fig. 4, (c)). At the same time, in the seeded mode, fitting data with a Gamma distribution does not well represent the measured distribution and gives the unlikely result of 58 modes (see Fig. 4, (a)), contradicting the spectral observation. A much better fit was provided by a Gaussian density function, with a number of modes close to one [25] (see Appendix C). This is also consistent with the statistical behaviour of the FERMI FEL-2 source conforming to that of a phase-locked laser.

We also analysed the dispersion values of the total intensity *I* given by $\zeta_{tot} = <\delta I^2>/<I>^2$, where $\delta I = I-<I>$. They were measured at the spectrometer as a function of the radiation bandwidth for both regimes of operation, Fig. 4 (b, d) and Appendix D. In the SASE mode, Fig. 4 (d), we observed the typical behaviour measured also at other SASE FEL sources [9, 10] with saturation at the narrower bandwidth values. We found that this saturation value was smaller (0.5) than the expected value of unity. Such behaviour was observed previously [10] and may be explained by the limited spectrometer resolving power (see Appendix E). In the seeded mode, we observed similar behaviour of the dispersion values (see Fig. 4 (b)), but the saturation value corresponding to a narrow bandwidth was much lower, about 0.07.

For coherent laser radiation we expect the $g^{(2)}$ function to be equal to one in the whole spatial and spectral range [18]. The residual contribution of 2%-7% at FERMI may be attributed to a combination of "incoherent" sources of noise and variations of "external" parameters, such as electron beam energy or trajectory. Even if stabilized by feedbacks, these parameters are subject to shot-to-shot variations affecting the properties of the emitted light. Such fluctuations do not affect the coherence of individual pulses, but may affect the average value in equation (1). Among genuine chaotic contributions, a residual contribution from microbunching instability may be mixing noise with the coherent energy modulation of the seed (see Appendix E). In a seeded FEL, the statistical fluctuations of the seed laser itself are translated to the XUV by the harmonic conversion process. Values of $g^{(2)}$ higher than one are often observed in optical lasers [26], caused by mixing of chaotic radiation or quantum noise with the coherent laser radiation.

A combination of HBT interferometry and spectral measurements allowed us to demonstrate that a seeded FEL is fundamentally different in its statistical properties from a SASE-based FEL. These measurements are a decisive step forward in understanding the basic properties of FELs. While SASE FELs behave statistically as chaotic sources, the seeded FERMI FEL is equivalent in its statistical properties to a coherent laser in the definition of Glauber. By performing high-order correlation analysis we show that a range of quantum optics experiments previously explored with optical fields can be performed with x-rays. For example, the Hong-Ou-Mandel effect [27], in which the quantum interference of indistinguishable photons is more intense than that of classical waves, should be observable with the light from FERMI. Coincidence detection by two detectors (second-order correlation measurements) may resolve the question raised recently in Ref. [28] on diffraction of stimulated emission with intense FEL light. The application of ideas and methods based on high-order coherence at x-ray energies is in its early stage of development [16,19], and the knowledge that second-order coherent FEL light is available permits the design and execution of new experiments, thus opening up this frontier. An open and intriguing question regards the statistical properties of self-seeded FELs [29, 30]. In contrast to an externally seeded FEL, this ideally first-order coherent source may show different second-order statistical properties.

**Methods**

*a. Correlation functions*

The normalized *first-order* correlation function is defined as

$$g^{(1)}(\mathbf{r}_1, \mathbf{r}_2) = \frac{\langle E^*(\mathbf{r}_1) E(\mathbf{r}_2) \rangle}{\langle \sqrt{I(\mathbf{r}_1)} \rangle \langle \sqrt{I(\mathbf{r}_2)} \rangle}, \tag{2}$$

where E(**r₁**) and E(**r₂**) are complex amplitudes of the wave field at different spatial positions measured simultaneously, and the brackets < … > denote averaging over a large ensemble of different radiation pulses. The first-order correlation function (2) represents the mutual intensity function [25].

The normalized *second-order* intensity correlation function $g^{(2)}(\mathbf{r}_1,\mathbf{r}_2)$ is defined by equation (1). An important quantity derived from the second-order intensity correlation function (1) is the contrast defined as

$$\zeta_2(\mathbf{r}) = g^{(2)}(\mathbf{r}, \mathbf{r}) - 1. \tag{3}$$

Chaotic sources with Gaussian statistics may be described by the following intensity correlation function [9,31]

$$g^{(2)}(\mathbf{r}_1, \mathbf{r}_2) = 1 + \zeta_2(D_\omega) |g^{(1)}(\mathbf{r}_1, \mathbf{r}_2)|^2. \tag{4}$$

Here $\zeta_2(D_\omega)$ is the contrast function defined in (3), which in the case of chaotic sources depends strongly on the radiation frequency bandwidth $D_\omega$ and $g^{(1)}(\mathbf{r}_1,\mathbf{r}_2)$ is the first-order correlation function (2). In the case of a chaotic pulsed beam, $\zeta_2(D_\omega)$ is determined by $\tau_c/T$, where $\tau_c = 2\pi/D_\omega$ is the coherence time and T is the pulse duration of the FEL radiation [31]. In this limit the number of longitudinal modes $M_t$ defined as $M_t = T/\tau_c$ is inversely proportional to the contrast function $\zeta_2(D_\omega)$. Notice also that for a perfect chaotic Gaussian source $g^{(2)}(\mathbf{r},\mathbf{r}) = 1 + \zeta_2(D_\omega)$ and does not depend on position **r**.

*b. Experimental details*

We performed the experiment at the beamline DiProI of the FERMI FEL-2 source with both seeded and SASE modes. The scheme of the experiment is shown in Fig. 1. The double cascade source FEL-2 tuned to the wavelength of 10.9 nm was used to generate seeded and SASE radiation. The radiation was focused with a Kirkpatrick-Baez optical system, and the intensity distribution was detected at a distance of 0.5 m from the focus. The beam divergence, after the refocusing optics, was about 1.0 mrad. An in-vacuum Andor Ikon CCD detector with 2048x2048 pixels of 13.5 μm x 13.5 μm size was used for intensity measurements of the direct beam. A small portion of the beam was partially diffracted by a grating to the shot-by-shot PRESTO spectrometer [32]. This diagnostic provides spectrum of each pulse simultaneously with the measurements at the DiProI end station. Several series of both direct beam and spectral images at different FEL parameters were recorded. Each series consisted of $10^4$ shots, measured with 10 Hz frequency.

The spectrometer is characterized by a resolving power of $1.8 \cdot 10^4$ at 10.9 nm [32], corresponding to the separation of two spectral lines of about $6 \cdot 10^{-4}$ nm. This implies that the spectrometer resolving power is sufficient to correctly measure the width of the seeded FEL pulses ($5 \cdot 10^{-3}$ nm). The SASE spectrum contains spikes resulting from the spectral superposition of uncorrelated temporal spikes separated in time. The duration of the temporal spikes can be derived from the SASE average spectral width ($6.5 \cdot 10^{-2}$ nm) and corresponds to about 6 fs. The finest structure in the spectral distribution depends on maximum temporal separation between the spikes. Assuming this separation to be of the order of 1 ps, i.e. the duration of the electron current (see Appendix E) we calculate the width of these finest structures in the spectrum to be about $4 \cdot 10^{-4}$ nm, which would require a resolving power of about $3 \cdot 10^{-4}$ to be observable. For this reason we found that the saturation value in Fig. 4(d) was smaller (0.5) than the expected value of unity. At the maximum spectral resolution of the spectrometer, it may still collect about two longitudinal modes which gives the contrast value in equation (3) of about 0.5.

**Acknowledgements**

We acknowledge E. Weckert for helpful discussions and support of the project. We thank the whole staff of FERMI for their support and excellent operation of the facility as well as for the fruitful discussions during this experiment; especially we are thankful to G. Penco and G. De Ninno from the FERMI facility. We acknowledge fruitful discussions with F. Kärtner and careful reading of the manuscript by Yu. N. Obukhov and T. Laarmann.

**Author contributions**

I.A.V., W.W., K.C.P., and M.K. conceived the experiment and coordinated the experimental efforts. L.G. planned FERMI FEL-2 operation. M.K., F.C., M.M. and E.P.

prepared the DiProI end-station for measurements. O.Yu.G., G.M., F.C., P.S., S.L., I.Z., M.D., M.D.`A, M.M., E.P., M.K., K.C.P., W.W., and I.A.V. performed the measurements. O.Yu.G. and P.S. analysed the data. O.Yu.G, I.A.V., L.G., and K.C.P. wrote the manuscript with contributions and improvements from all authors. All authors read and discussed the manuscript.

**Competing interests:** The authors declare no competing financial interests.

**APPENDIX A: Additional experimental results**

Projections of the averaged intensity distribution (black lines) in the vertical and horizontal directions for the seeded and SASE modes of operation are shown in Fig. 5. It can be well seen from these figures that intensity distribution is significantly narrower in the seeded regime. Intensities of individual pulses (blue lines) are similar in the shape to the averaged intensity that suggests that in the transverse direction we have predominantly a single mode.

Intensity correlation functions in the horizontal direction for the seeded and SASE modes of operation are shown in Fig. 6. They exhibit similar behavior to the vertical direction.

Cross sections of the intensity correlation functions in the vertical and horizontal directions along the two diagonals (red and white) in Fig. 2 (a, c) and Fig. 6 for the seeded and SASE modes are presented in Fig. 7 and Fig. 8. The values below one (anticorrelation) in $g^{(2)}$-function (see Fig. 8) may appear due to positional jitter [10].

Intensity correlation functions in frequency domain measured at the spectrometer are presented in Fig. 9 and display very pronounced features typical for the jitter-affected intensity distribution [10] in the seeded regime (Fig.9 (a)). It suggests that significant jitter of the carrier wavelength was present during our measurements.

**APPENDIX B: Sorting of spectral modes**

In order to analyze the spectral dependence of the $g^{(2)}$ function in seeded mode of operation, we implemented a sorting procedure. Spectra of individual pulses were fitted with multiple Gaussian distributions and sorted versus the relative percentage of energy in the most intense mode using the following procedure. Pulses containing multiple longitudinal modes generally contain several peaks in the spectral profile. In order to estimate the number and contribution of each mode, the spectral distribution of each pulse was fitted with several Gaussian distributions (multiple peak Gaussian distribution). The spectral profile of each pulse was first fitted with the one Gaussian function, then with two, and so on. The number of distributions was not increased further if one of the two conditions was fulfilled. First, when the unadjusted coefficient of determination $R^2$ reaches one with the value smaller than $10^{-4}$; or second, when the area under the most intense peak, that correspond to the power of the main mode, is decreased by more than 10% of the total area under the spectrum. Last criterion avoids overfitting of the data. Additionally, the maximum width (FWHM) of each Gaussian spectral distribution has an upper limit of 0.2 $fs^{-1}$ in order to avoid appearance of peaks wider than the whole pulse spectrum. By that we selected $10^3$ pulses with the largest contribution of the main mode and $10^3$ pulses with the smallest contribution.

**APPENDIX C: Intensity distribution for chaotic and laser sources**

In the case of chaotic source obeying Gaussian statistics, the probability $p(I)$ that the total intensity of the pulse takes value $I$, follows Gamma distribution [25,33]

$$p(I) = \frac{M^M}{\Gamma(M)} \left(\frac{I}{\langle I \rangle}\right)^{M-1} exp\left(-M \frac{I}{\langle I \rangle}\right), \qquad (1)$$

where M is the number of degrees of freedom, or modes, and $\langle I \rangle$ is an average intensity of the pulse.

In the case of laser radiation, probability distribution of the total intensity is mostly explained by various noise effects and obeys a normal distribution [25]

$$p(I) = \frac{1}{\sqrt{2\pi\sigma^2}} exp\left(-\frac{(I-\langle I \rangle)^2}{2\sigma^2}\right), \qquad (2)$$

where σ is the width of the distribution.

**APPENDIX D: Analysis of intensity dispersion as a function of the bandwidth**

We analysed dispersion of the total intensity $I_{tot}$ by using the two-dimensional intensity distribution on the spectrometer detector as a function of the radiation bandwidth for both regimes of operation. Typical images obtained at the spectrometer in the seeded and SASE modes of operation are shown in Fig. 10. By choosing different bandwidths around the carrier frequency and integrating intensity in the considered region, the value of the total intensity in the selected bandwidth for each pulse is determined. The relative dispersion is than calculated as

$$\zeta_{tot} = \frac{\langle (I-\langle I \rangle)^2 \rangle}{\langle I \rangle^2}. \qquad (3)$$

**APPENDIX E: FERMI operation**

FERMI FEL-2 employs a double cascade of high gain harmonic generation (HGHG). The HGHG scheme [34,35] consists in preparing the electron beam phase space in a first undulator, called modulator, where the interaction with an external laser, the seed, induces a controlled, periodic modulation, in the beam energy distribution. The beam then traverses a "dispersive section", which converts the energy modulation into a density modulation. The higher order harmonic components of this modulation retain the phase and amplitude properties of the seed. The "density" modulated beam is then injected into an FEL amplifier, resonant to the desired higher order harmonic, where the startup of the amplification process is enhanced by the presence of the modulation. The modulation depth may be tuned by varying the seed intensity or the dispersion, in order to reach saturation and efficient energy extraction at the end of the amplifier. This HGHG scheme is implemented in FERMI FEL-1 to generate fully coherent radiation pulses in the VUV spectral range, from 100 nm to 20 nm. The amplitude of the energy modulation necessary to initiate the HGHG process grows with the order of the harmonic conversion. The induced energy dispersion has a detrimental effect on the amplification in the final radiator at higher harmonic orders. For this reason in FERMI FEL-2 the harmonic multiplication process is repeated twice. In a first HGHG stage an intermediate harmonic is produced, that is then used as a seed in a second HGHG stage. This double conversion is achieved with the fresh bunch injection technique [34,35] where a delay line slows down the electron bunch to accommodate the seed from the first stage on a fresh portion of the electron current, allowing the second harmonic conversion on electrons with low energy spread. FEL-2 is designed and operates in the spectral range from 20 nm to 4 nm, at the upper edge of the water window. This experiment was done with the seed initiating the cascade produced by a Ti:Sa oscillator, amplified in a regenerative amplifier and converted to the third harmonic at 261.5 nm. The first stage of FEL-2 was operated at the $6^{th}$ harmonic of the seed while the second stage was operated at the $4^{th}$ harmonic of the first stage, for a total harmonic conversion factor of 24, corresponding to 10.9 nm of final output wavelength. A seed energy of 15 µJ was sufficient to induce a coherent modulation in the electron current at the entrance of the amplifier three orders of magnitude larger than the shot noise background. This estimate suggests that the SASE background power was $10^6$ times lower than the seeded FEL power. In terms of energy we have to consider that the pulse duration in seeded mode is expected to be of the order of 40-50 fs [36] in the operating conditions of the experiment, while in SASE mode most of the beam, about 1 ps long, may contribute to emission. We would therefore expect to have an energy contrast between the seeded coherent signal and the SASE Gaussian noise of approximately $10^4$ - $10^5$.

This estimate does not include the effects of macroscopic or microscopic modulations of the beam current that could enhance the SASE emission or directly induce modulations of chaotic nature in the emitted radiation. The main player in this respect is the microbunching instability amplifying existing modulations in the beam that are then mixed with the energy modulation induced by the seed and may determine an observable effect on the FEL spectrum [37], as the formation of SASE like structures. The FEL operating conditions were set in order to minimize these effects. However, a diagnostic based on HBT interferometry is very promising in distinguishing the presence of microbunching instability and may contribute to the mitigation of microbunching instability during the FEL tuning process.

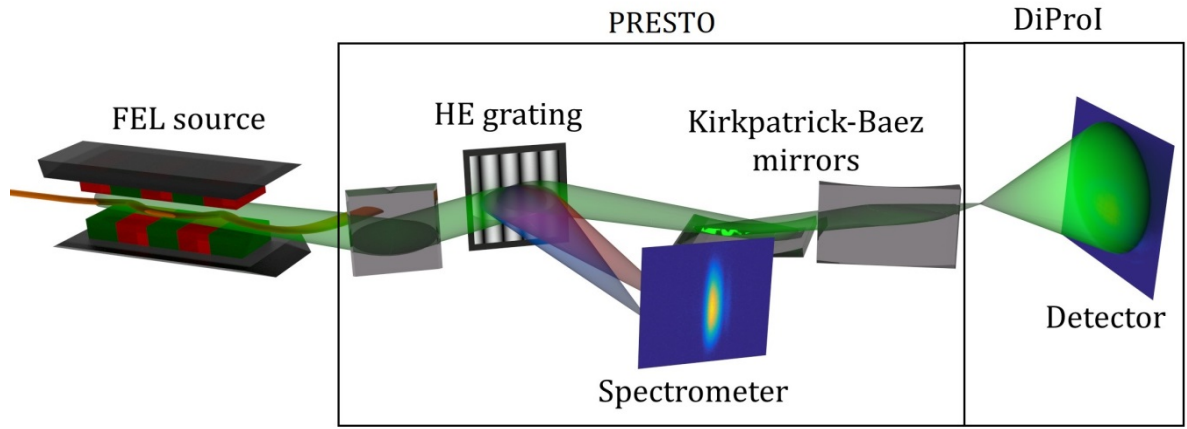

FIG. 1: **Schematic layout of the experiment.** Radiation generated in the undulators is focused by KB mirrors and the detector is installed out of focus to observe the direct beam. Radiation from each pulse is partially diffracted by a grating to the spectrometer detector to observe on-line pulse spectra.

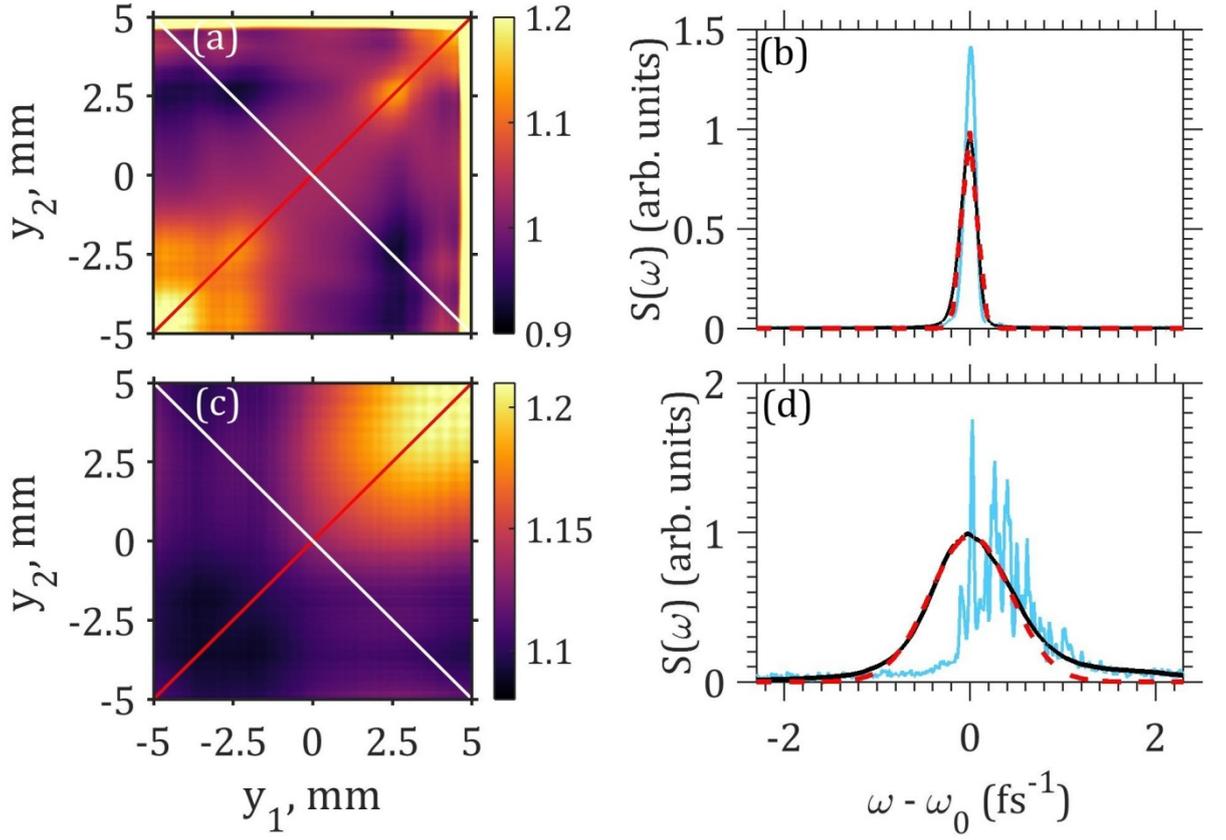

FIG. 2: **Seeded and SASE mode correlation functions and spectra.** Intensity correlation functions for seeded (a) and SASE (c) regimes of operation. Spectral structure of an individual pulse (blue line), average spectrum (black line) and Gaussian fit (red dashed line) for seeded (b) and SASE (d) regimes of operation.

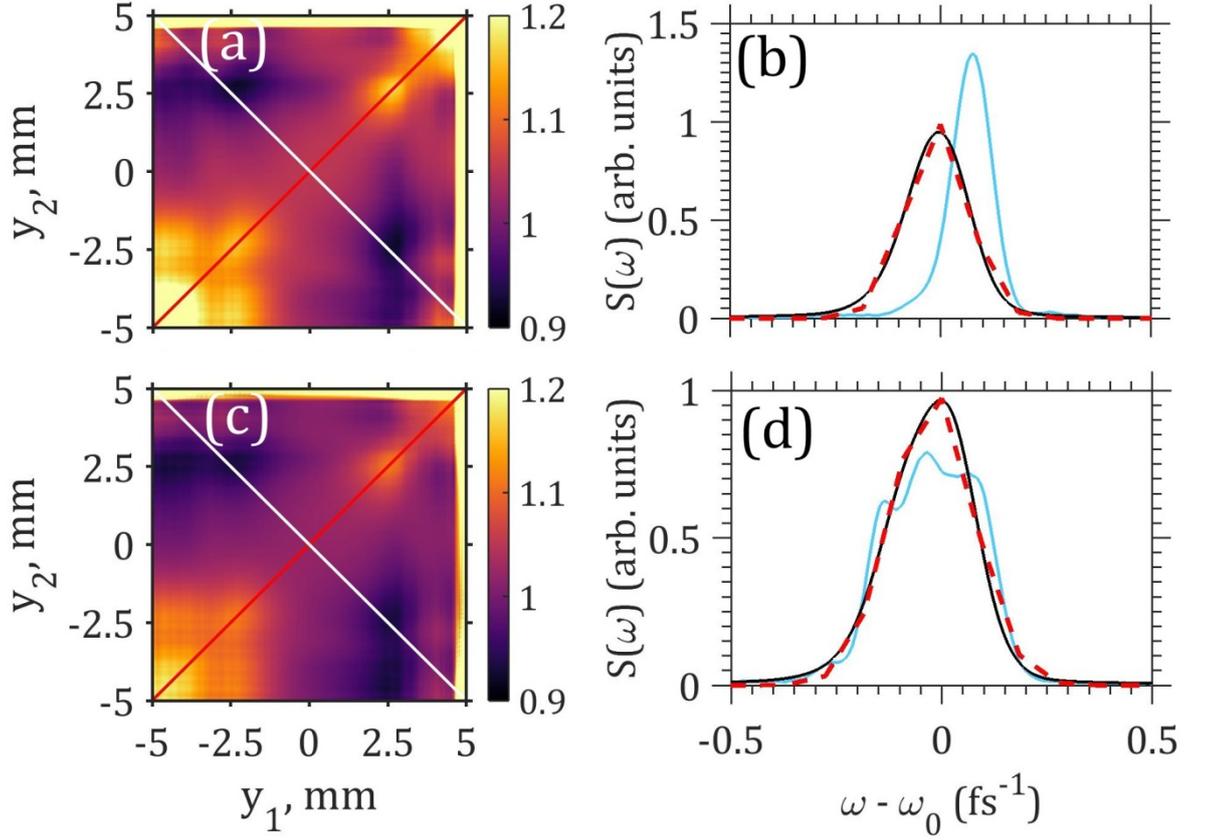

FIG. 3: **Seeded mode single and multiple pulses analysis.** Intensity correlation functions (a,c) and spectral structure (b,d) for the sorted pulses with the lowest (a,b) and highest (c,d) amount of modes. In (b, d) representative single pulses are shown by blue lines, an average over $10^3$ pulses is shown by black lines, and Gaussian fit by red dashed lines. Note the different scale in frequency ω in comparison with Fig. 2 (b,d).

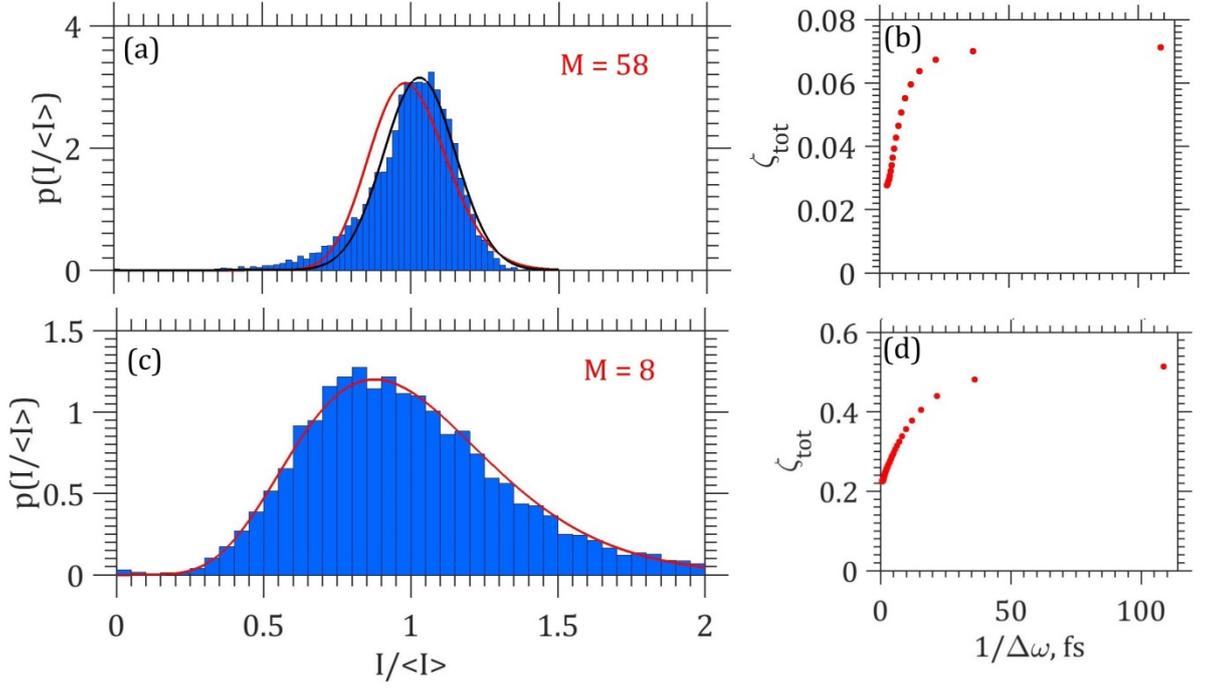

FIG. 4: **Seeded and SASE mode histograms and dispersion values of intensity.** (a,c) Histograms of intensity for seeded (a) and SASE (c) regimes. The red line represents a fit by a Gamma function and black line represents a fit by a Gaussian function. Number of modes determined from the Gamma distribution is indicated by M. (b,d) Contrast behaviour as a function of inverse bandwidth for seeded (b) and SASE (d) regimes of operation.

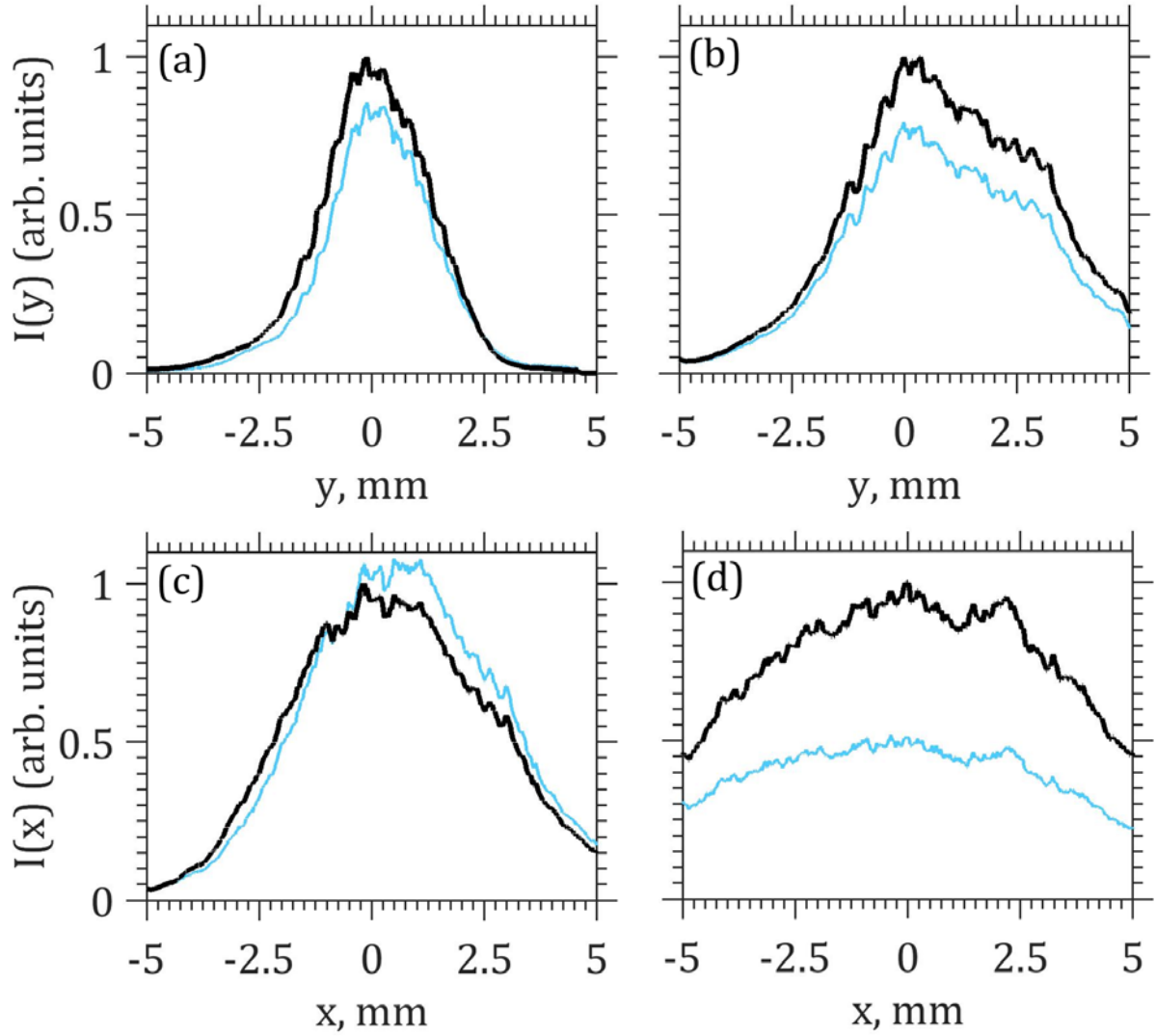

FIG. 5: Projection of the averaged intensity distribution (black lines) in the vertical (a,b) and horizontal (c,d) directions for the seeded (a,c) and SASE (b,d) modes of operation. Blue lines are individual pulses.

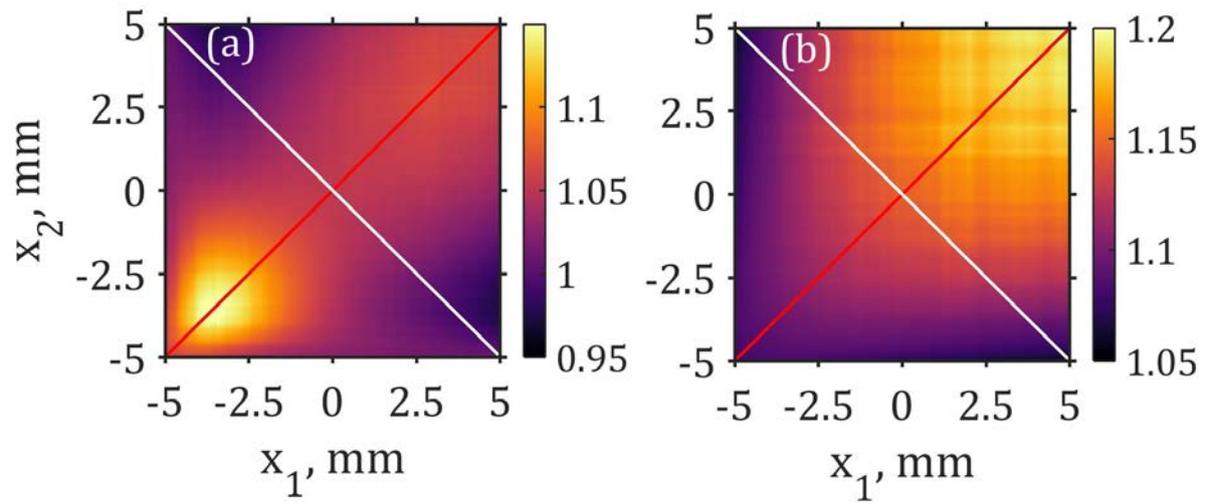

FIG. 6: Intensity correlation functions in the horizontal direction in the seeded (a) and SASE (b) modes of operation.

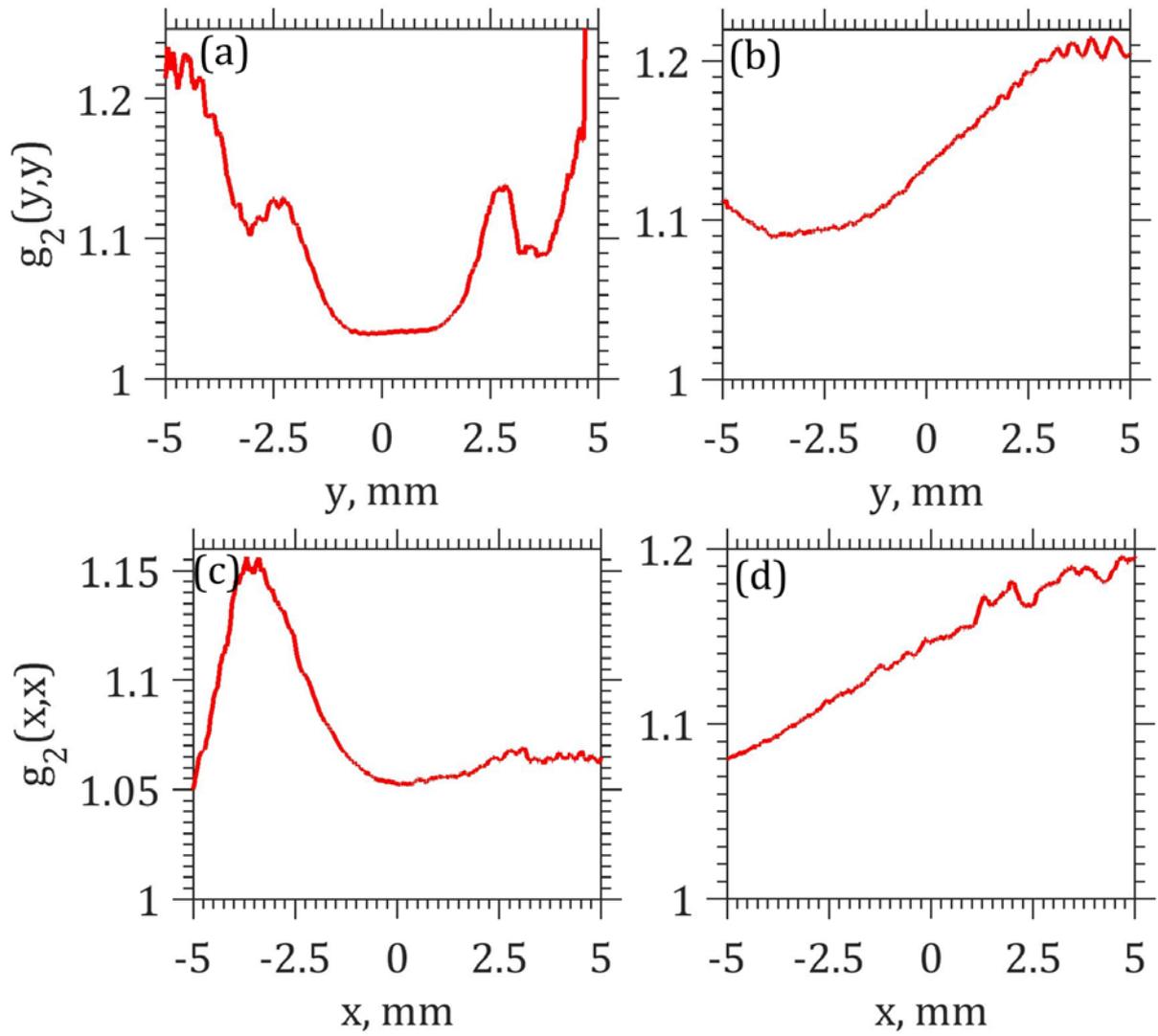

**FIG. 7: Cross sections of the intensity correlation function along the diagonal (red line) in Fig. 2 in the vertical direction (a, b) and in Fig. 6 in the horizontal direction (c, d) in the seeded (a, c) and SASE (b, d) regimes of operation.**

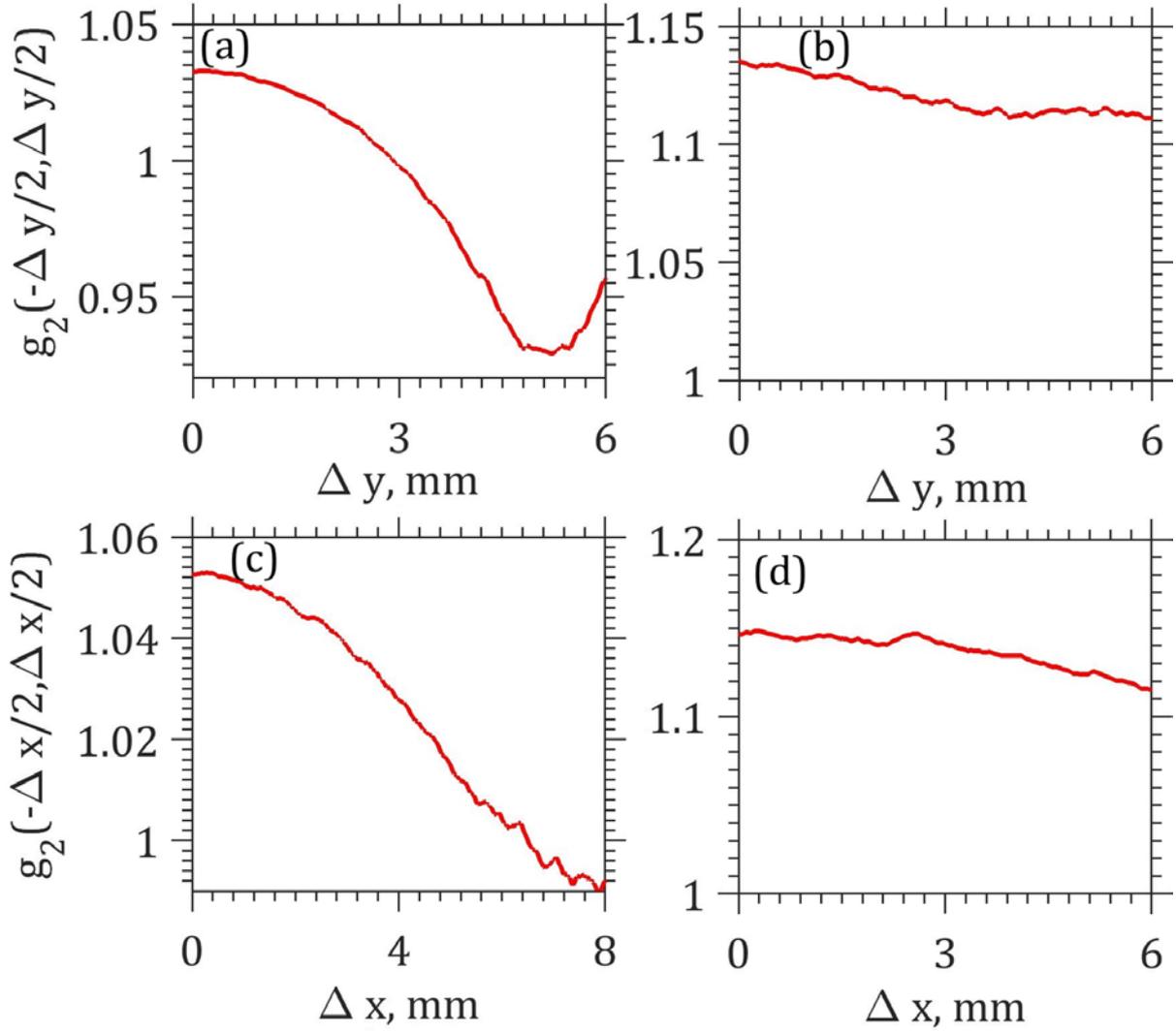

**FIG. 8:** Cross sections of the intensity correlation function along the diagonal (white line) in Fig. 2 in the vertical direction (a, b) and in Fig. 6 in the horizontal direction (c, d) in the seeded (a, c) and SASE (b, d) regimes of operation.

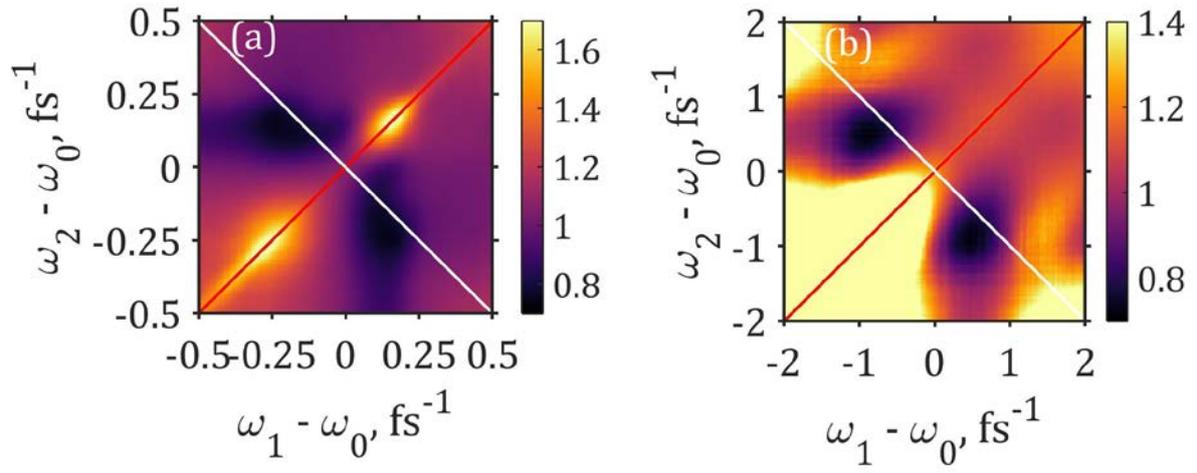

FIG. 9: Intensity correlation functions in frequency domain measured at the spectrometer for seeded (a) and SASE (b) regimes of operation.

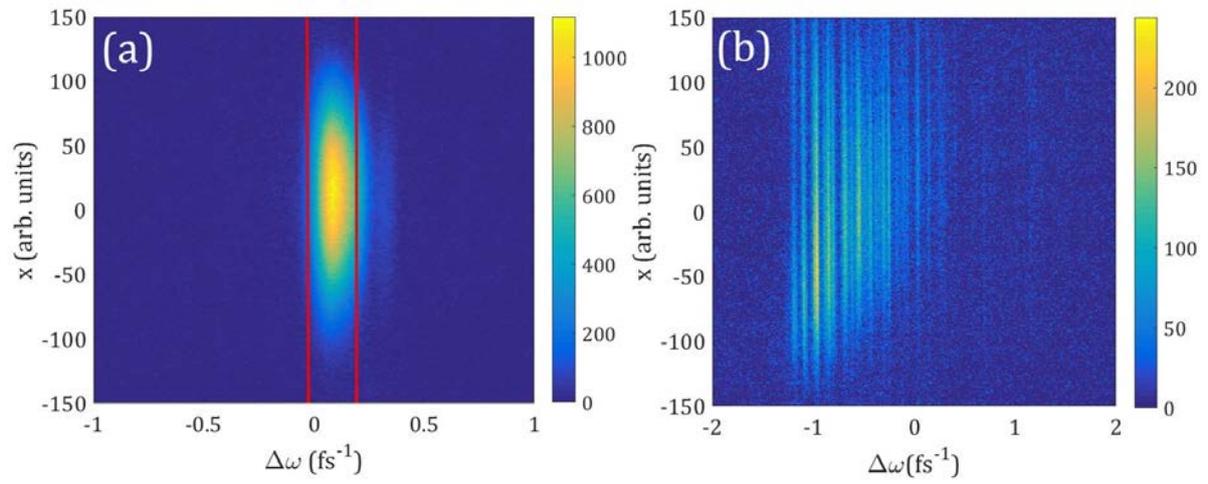

FIG. 10: Typical images on the spectrometer detector in seeded (a) and SASE (b) modes of operation. Red lines show an example of a chosen bandwidth for analysis. Note twice wider spectral range in SASE mode.